\documentclass[manuscript]{aastex}

\usepackage{color}

\newcommand{\kms}{km~s$^{-1}$}
\newcommand{\degree}{\ensuremath{^\circ}}

\newcommand{\suny}{Sun-y}
\newcommand{\sunx}{Sun-x}
\shorttitle{Automatic detection of solar EUV dimmings}
\shortauthors{Alipour, Safari, Innes }

\begin{document}


\title{AN AUTOMATIC METHOD FOR EXTREME-ULTRAVIOLET DIMMINGS ASSOCIATED WITH
SMALL-SCALE ERUPTION}

\author{N. Alipour\altaffilmark{1}, H. Safari\altaffilmark{1}, and D. E. Innes\altaffilmark{2} }
\affil{$^1$Department of Physics, University of Zanjan, P. O. Box 45195-313, Zanjan, Iran\\
$^2$Max-Planck Institut f\"{u}r Sonnensystemforschung, 37191
Katlenburg-Lindau, Germany }
%\email{safari@znu.ac.ir}
%
%\and
%
%\author{H. Safari}
%\affil{Department of Physics, Zanjan University, P. O. Box
%45195-313, Zanjan, Iran} \email{safari@znu.ac.ir}
%\author{C. D. Biemesderfer\altaffilmark{4,5}}
%\affil{National Optical Astronomy Observatories, Tucson, AZ 85719}
%\and
%\author{R. J. Hanisch\altaffilmark{5}}
%\affil{Space Telescope Science Institute, Baltimore, MD 21218}
%% are identified by the \altaffilmark after each name. Specify alternate
%\altaffiltext{1}{Visiting Astronomer, Cerro Tololo Inter-American Observatory.
% CTIO is operated by AURA, Inc.\ under contract to the National Science Foundation.}
%\altaffiltext{2}{Society of Fellows, Harvard University.}
%\altaffiltext{3}{present address: Center for Astrophysics, 60 Garden Street, Cambridge, MA 02138}
%\altaffiltext{4}{Visiting Programmer, Space Telescope Science Institute}
%\altaffiltext{5}{Patron, Alonso's Bar and Grill}

\begin{abstract}\textbf{Small-scale extreme ultraviolet (EUV) dimming often surrounds sites of energy release in the quiet Sun.
This paper describes a method for the
  automatic detection of these small-scale EUV dimmings}
using a feature based classifier. The method is demonstrated
using sequences of 171\AA\ images taken by STEREO/EUVI on 13 June
2007 and by SDO/AIA on 27 August 2010. The feature identification
relies on recognizing structure in sequences of space-time
171\AA\ images using the Zernike moments of the images. The
Zernike moments space-time slices with events and non-events are
distinctive enough to be separated using a Support Vector Machine
(SVM) classifier. The SVM is trained using 150 event and 700
non-event space-time slices. We find a total of 1217 events in
the EUVI images and 2064 events in the AIA images on the days
studied. Most of the events are found between latitudes
-35\degree\ and +35\degree. The sizes and expansion speeds of central
dimming regions are
 extracted using a region grow algorithm. The histograms of the
  sizes in both EUVI and AIA follow a steep power law with slope about -5.
   The AIA slope extends to smaller sizes before turning over.
    The mean velocity of 1325 dimming regions seen by AIA is found to be about 14~\kms.

\end{abstract}
\vspace{17cm}
\newpage
\clearpage
\section{Introduction} {Extreme ultraviolet (EUV) dimming surrounding small bursts of EUV brightening in the quiet Sun were first noticed by Innes et al. (2009) in 171\AA\
images taken by the Extreme UltraViolet Imager
(EUVI) on STEREO. The dimmings are usually more extended and last longer than the brightening (Podladchikova et al. 2010; Innes et al 2010). They seem to be related chromospheric eruptions at the junctions of supergranular cells (Innes et al. 2009).
In 171\AA\ solar disk images small-scale eruptions have similar
characteristics to face-on CMEs but on a smaller scale.
Apart from microflare brightening and dimming of the surrounding corona,
several events have wave-like features propagating from the
eruption site.}

Podladchikova et al. (2010) focused on the small-scale diffuse
coronal fronts and their associated dimmings for a couple of
events. The coronal wave and dimming properties were extracted
using a semi-automatic algorithm adapted from their method
developed for CMEs (Podladchikova \& Berghmans 2005).
They show that waves develop from micro-flaring sites and
propagate up to a distance of 40,000~km in 20~min. The dimming
region is two orders of magnitude smaller than for large-scale
events. The waves associated with the micro-eruptions had a
velocity of 17~\kms, and were diagnosed as slow mode, propagating
almost perpendicular to the background magnetic field.

Innes et al. (2009) estimated a rate of 1400 events per
day over the whole Sun by picking out events by eye in a period
of 24 hours. Now the Solar Dynamics Observatory/Atmospheric Imaging Assembly
(SDO/AIA) is providing full Sun images through ten UV and EUV
filters with a cadence of about one image every second. To
optimize statistical analyses of various kinds of solar phenomena,
it is necessary to develop automatic detection
techniques. Aschwanden (2010) gives an extensive overview of
solar image processing techniques that are used in automated
feature detection algorithms. One of the modules in the Computer
Vision Center for SDO is designed to automatically detect and
extract coronal dimmings from AIA images (Attrill \& Wills-Davey
2010).

 Here, we give an automatic method to detect
 {small-scale dimmings} in the 171\AA\ STEREO/EUVI and SDO/AIA images using
Zernike moments and the Support Vector Machine (SVM) classifier.
The method exploits the characteristics of the Zernike moments in
order to represent event/non-event patterns. The Zernike moments
are invariant to rotation, scaling, and translation, and this
allows the proposed method to detect events even if they are
translated, scaled or rotated. First the
 Zernike moments are extracted for a selection of events and non-events, and fed into the SVM
 as training data sets so that it can
successfully identify events/non-events in subsequent 171\AA\
image sequences.

 The paper layout is as
follows: data reduction is discussed in Section \ref{data};
Zernike moments are explained in Section 3; the Support Vector
Machine is discussed in Section \ref{svm}; and the results and
conclusions are presented in Section \ref{con}.

\section{Data Reduction }\label{data}
EUVI on the STEREO spacecraft takes images of the Sun through
four different EUV filters (171\AA, 195\AA, 284\AA\, and 304\AA)
(Howard et al. 2008). In the present work we have restricted ourselves
to 171\AA\ images with a time cadence of 2.5~min and a pixel size
1.6\arcsec, taken on 13 June 2007. First the recorded images were
calibrated using the SolarSoft routine secchi$\_$prep.pro.

SDO is designed to study the solar atmosphere on small scales of
space and time and in many wavelengths simultaneously
(\url{http://sdo.gsfc.nasa.gov/mission/instruments.php}). For
this study we have used AIA level 1.5 171\AA\ data from the series
aia\_test.synoptic2 recorded on 27 August 2010. These are binned
data (1024$\times$1024) with a pixel size 2.4\arcsec\ and an
image interval of 90~s. To keep the image and pixel sizes
approximately the same in both datasets the SDO images were
rebinned to 2048$\times$2048 pixels.

\begin{figure}
\centering
\includegraphics[width=\linewidth]{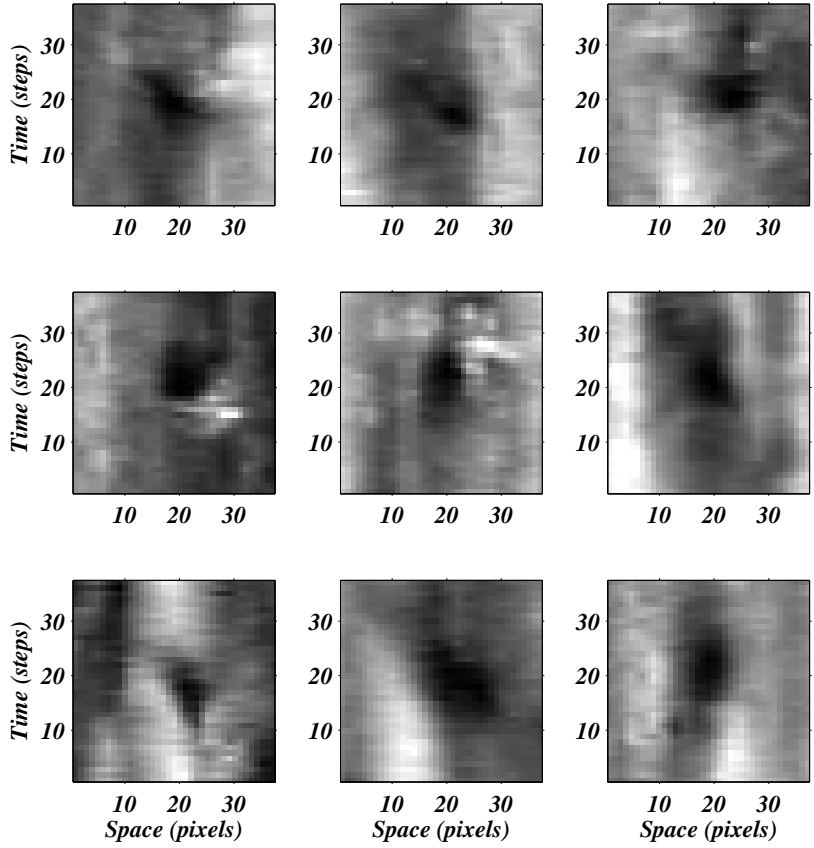}
       \caption{Examples of events shown as space-time slices (SDO/AIA). The
       frames are 37 pixels$\times$37 time steps.
       Each pixel is 1.2\arcsec\ across and each time step is 90~s. }
            \label{fig1}
\end{figure}

\begin{figure}
\centering
\includegraphics[width=\linewidth]{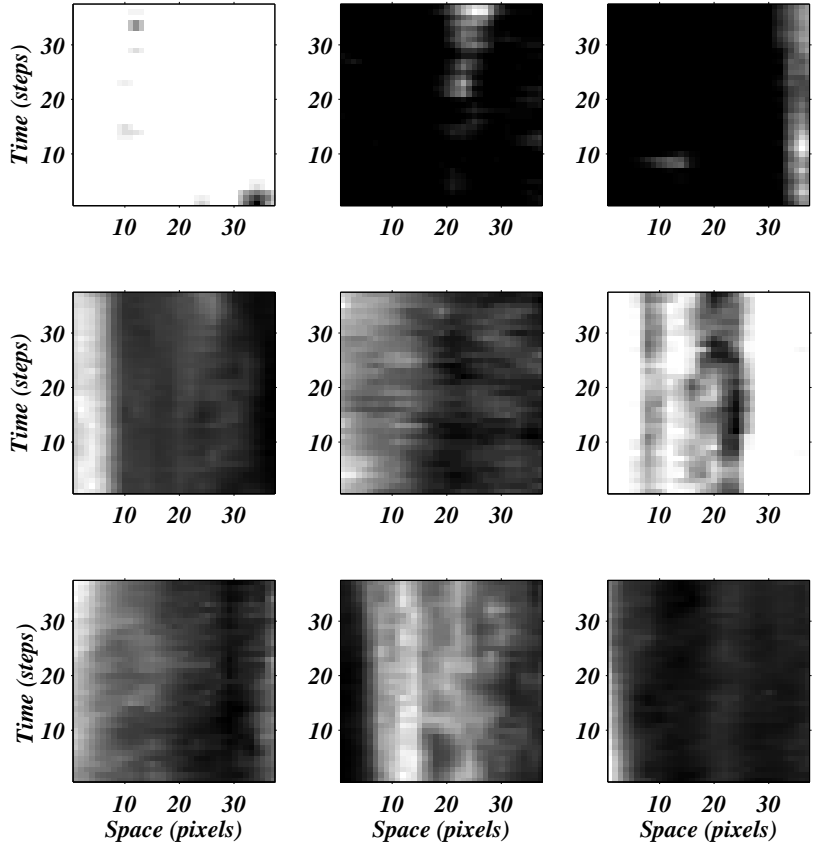}
       \caption{Examples of non-events shown as space-time slices (SDO/AIA). The
       frames are 37 pixels$\times$37 time steps.
       Each pixel is 1.2\arcsec\ across and each time step is 90~s. }
            \label{fig2}
\end{figure}

\begin{figure}
 \centering
\includegraphics[width=12cm]{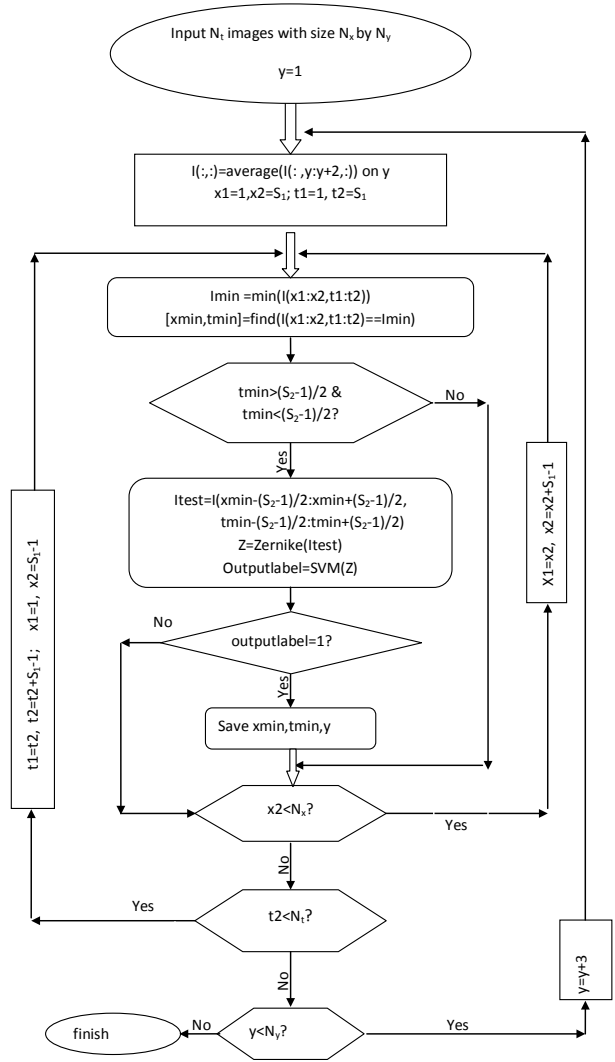}
\caption{A flowchart representation for the automatic detection
scheme for  {small dimmings}. The parameters used for EUVI were
$S_1=13$ and $S_2=31$, and for AIA $S_1=15$ and $S_2=37$. }
            \label{fig3}
\end{figure}

Innes et al. (2009)
     show that events that have expanding dimming regions produce
      distinctive diagonal structures in space-time EUV images
       (Figures~\ref{fig1} and \ref{fig2}).
      Our automatic procedure therefore systematically scans through
       space-time blocks of the data.
 A flowchart of the algorithm is shown in Figure~\ref{fig3}.
First all $N_t$ 171\AA\ images of size $N_x \times N_y$ from each
data set were calibrated and de-rotated to the start time of the
data series. Then following Innes et al. (2009) space-time slices
through the data were made by averaging 3 consecutive pixels
along the \suny\ direction to give a \sunx\ versus time image for
every third \suny\ position of
size $N_x \times N_t$. Then for each space-time image, a small
region, starting from $x1 = 1, t1 = 1$ with the size $\Delta
x=S_1$ and $\Delta t=S_1$ is extracted. The location, $x_{min}$,
and time, $t_{min}$, of the minimum intensity inside this small
region is determined. This gives the location of the deepest
dimming in the region. In step 3, a larger region is selected
around the minimum intensity position ($x_{min}$, $t_{min}$)
with size $\Delta x = S_2$ and $\Delta t = S_2$. The Zernike
moments, $Z_{pq}$, of this section of the image are computed. The
magnitudes of the moments are fed to the SVM classifier. The code
picks up a label 1 for an event class and 2 for a non-event
class. If it is an event, the location $x_{min}$ and time
$t_{min}$ are saved. Then the small box is moved first in space
until the end of the grid is reached and then in time. By
repeating this for each of the $N_y/3$ slices, all parts of the
data are investigated.

Neighbouring $y$-slices pick-up different parts of the same
event.  {Events found in neighbouring space-time slices are
grouped together so that the expansion speed and dimming size can
be measured in both the $x$ and $y$ directions.}

  \section{Zernike moments}\label{zernike}

  In the past decades, various moment functions (Legendre, Zernike,
  etc) due to their abilities to represent the image features have
  been proposed for describing images (Hu 1962). The Zernike
  polynomials form a complete set of orthogonal polynomials. The
  orthogonal property enables the contribution of each moment to be
  unique and independent of the information in an image. Therefore
  the Zernike moments, $Z_{pq}$, for each feature are unique.
  Space-time images of a slice on the Sun would look like simple
  vertical stripes if there is no activity. If there is a
  brightening or dimming with no flows the stripes change intensity
  but keep their basic vertical structure. When an  {eruption} occurs
  and the region expands, diagonal structures appear crossing the
  background vertical stripes. This change in structure is
  reflected in the pattern of the Zernike moments, and the high
  Zernike coefficients, $Z_{pq}$, appear in blocks. If we select a
  small box on the dimming region and then compute the moments the
  block structure in $Z_{pq}$ is not as clear as for a slightly
  large region that takes in the surroundings as well as the
  dimming region because it is the structure change that is
  registered in the moments.

\begin{figure}
 %\special{psfile=fig/fig4.eps vscale=30 hscale=28 hoffset=15
 % voffset=-200}
 % \vspace{6cm}
  \centering
\includegraphics[width=\linewidth]{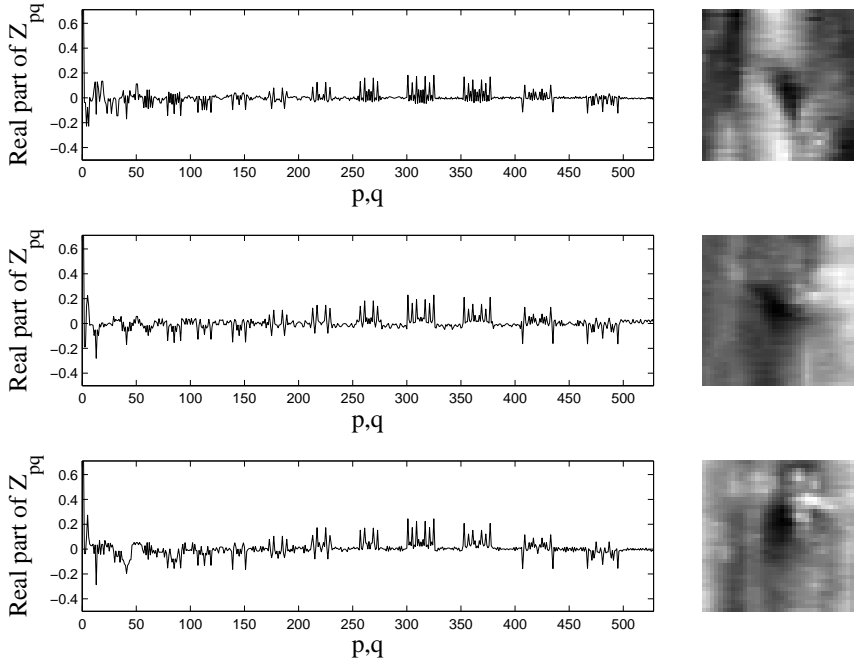}
      \caption{ Real part of Zernike moment, $Z_{pq}$,
 of three events.
       }
            \label{fig4}
\end{figure}

\begin{figure}
 %\special{psfile=fig/fig5.eps vscale=30 hscale=28 hoffset=15
 % voffset=-200}
 % \vspace{6cm}
   \centering
\includegraphics[width=\linewidth]{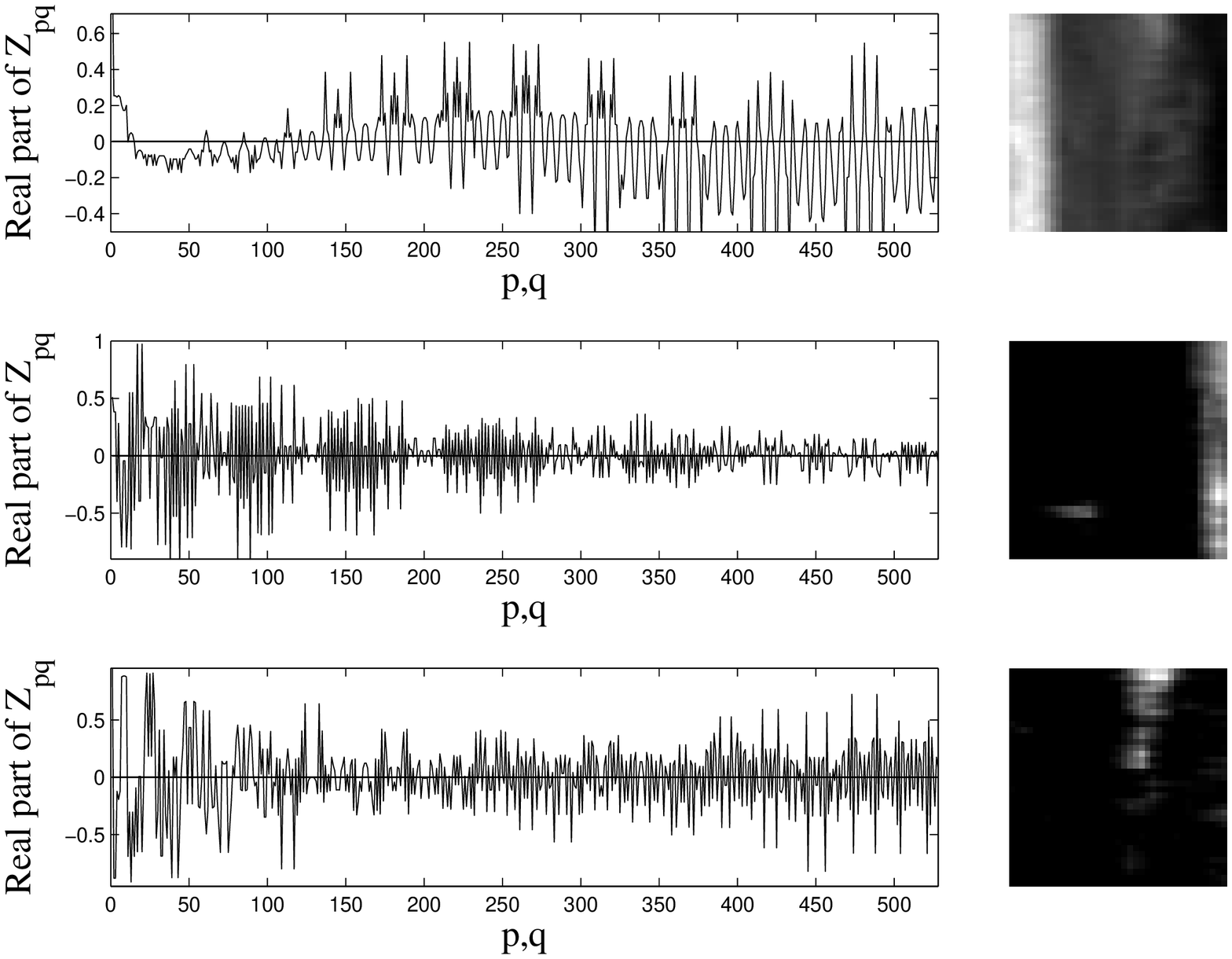}
      \caption{Real part of Zernike moment of the three non-events.
        }
            \label{fig5}
\end{figure}

  Using a Matlab code, we calculated the Zernike moments of order
  up to $p=31$. The repetition number, $q$, satisfied $p-|q|$ is
  even. This gives a $(q,p)$ set labelled from 1 to 528.
  The Zernike moments of three events and three non-events are shown in
  Figures~\ref{fig4} and \ref{fig5}, respectively. As shown in the
  figures, the real part of the Zernike moments are clearly
  different for event and non-events features. The events have a
  well-defined block structure. For non-events features, we see some
  block structure but it is different from the event type of block.
  These differences gives us confidence in applying a Support Vector
  Machine (SVM) classifier to identify events in space-time slices.

\section{Support Vector Machine}\label{svm}
The support vector machine (SVM) is a generation learning system
based on statistical learning theory. It was successfully applied
to text categorization, image classification, etc. SVM is
basically defined for two class problems. It finds the optimal
hyperplane which maximizes the distance between the
nearest examples of both classes. It is a supervised learning
method based on kernel functions (Burges 1998). We use the least
square SVMlab Toolbox classifier in the Matlab environment with
the Gaussian Radial Basis Function as the SVM kernal (Gunn 1997).

To prepare (train) the network we computed the Zernike moments of
150 events and 700 non-events. Once the SVM had been
trained it could be used to classify the space-time images and
find events.

\section{Results and Conclusions}\label{con}
\subsection{Results}
\begin{figure}
%% \special{psfile=sunevents.ps vscale=40 hscale=36 hoffset=35
%% voffset=-280}
%% \vspace{8.1cm}
\centering
\includegraphics[width=\linewidth]{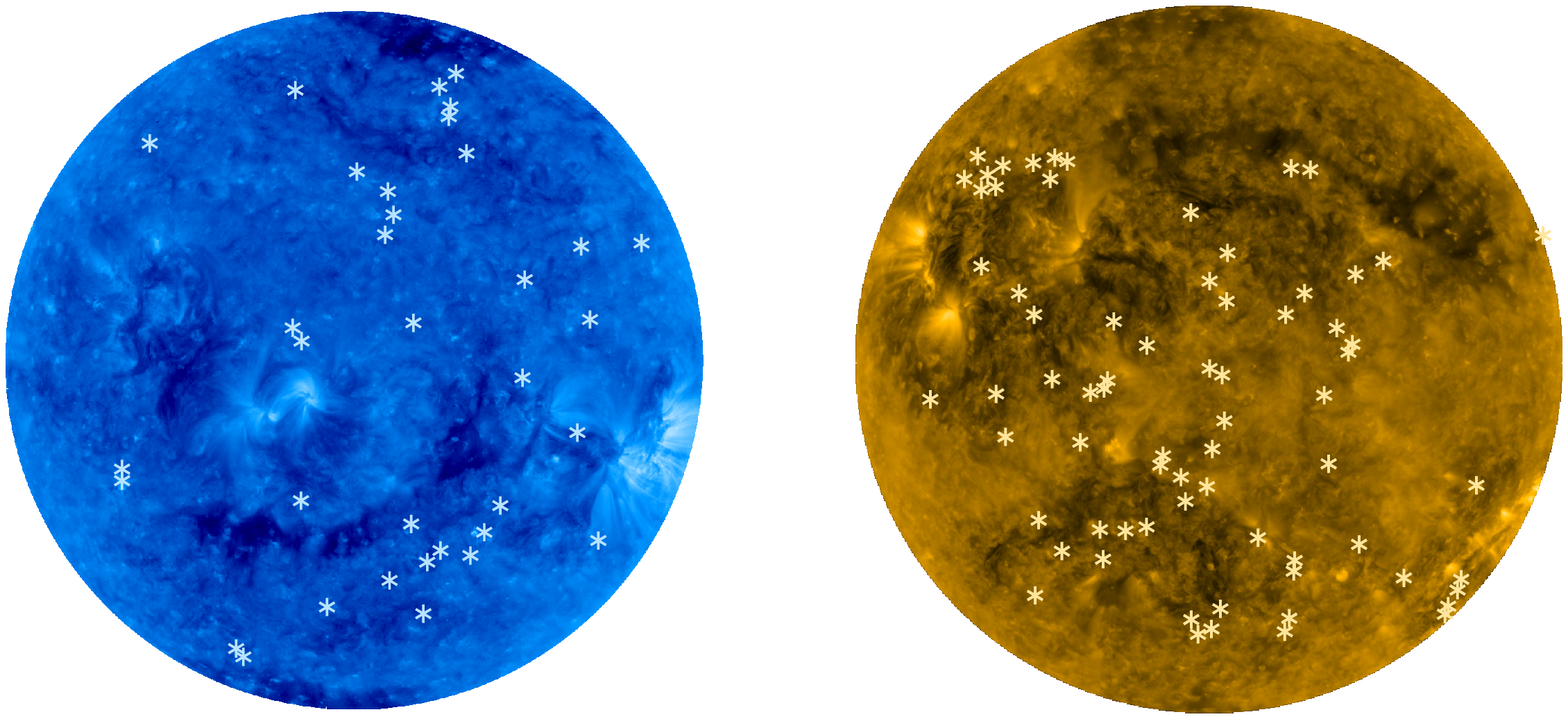}
       \caption{Positions of events detected in 20 min by EUVI (left) and AIA.}
            \label{fig7}
\end{figure}

The automatic detection scheme  {for small-scale EUV dimmings}
has been applied to full Sun STEREO/EUVI and SDO/AIA 171\AA\
data. On 13 June 2007, 1217 events were detected from the EUVI
images.
 This is approximately twice the rate found by Innes et al. (2009) by eye for the same dataset.
In the AIA images, 2064 events were detected on 27 August 2010.
AIA is therefore giving another factor two in event rate.
Figure~\ref{fig7} shows the positions of events found during
20~min of EUVI and AIA images.

\begin{figure}
% \special{psfile=fig/fig9.eps vscale=40 hscale=36 hoffset=-5
% voffset=-190}
% \vspace{4.5cm}
    \centering
\includegraphics[width=\linewidth]{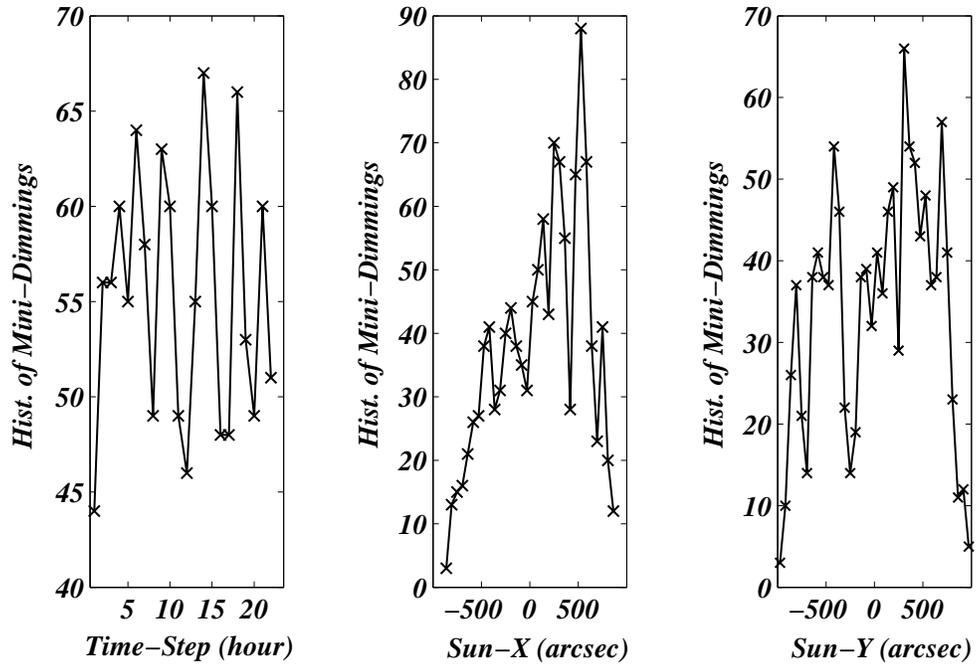}
       \caption{Number of events per hour and per 50\arcsec\ detected in STEREO/EUVI images on 13 June
       2007 plotted versus time (left)
        \sunx\ (middle), and \suny\ (right).
            }
            \label{fig8}
\end{figure}

The event numbers from EUVI are plotted versus time, \sunx, and
\suny\ in Figure~\ref{fig8}. We see that approximately 60 events
were detected every hour. On average, there are 45 events per
50\arcsec\ in both \sunx\ and \suny. The numbers for AIA were
double the EUVI numbers: 115 events per hour and 100 events for
every 50\arcsec\ in space. For both EUVI and AIA, the events are
mostly found in the central part of the disk (between +35 degrees
and -35 degrees). They are most easily identified against the
quiet Sun because in active regions they disappear into the
general background activity and off-limb there is too much
line-of-sight confusion (Innes et al. 2009).

\begin{figure}
 %\special{psfile=fig/fig6.eps vscale=50 hscale=66 hoffset=-45
 % voffset=-280}
 % \vspace{7cm}
    \centering
\includegraphics[width=\linewidth]{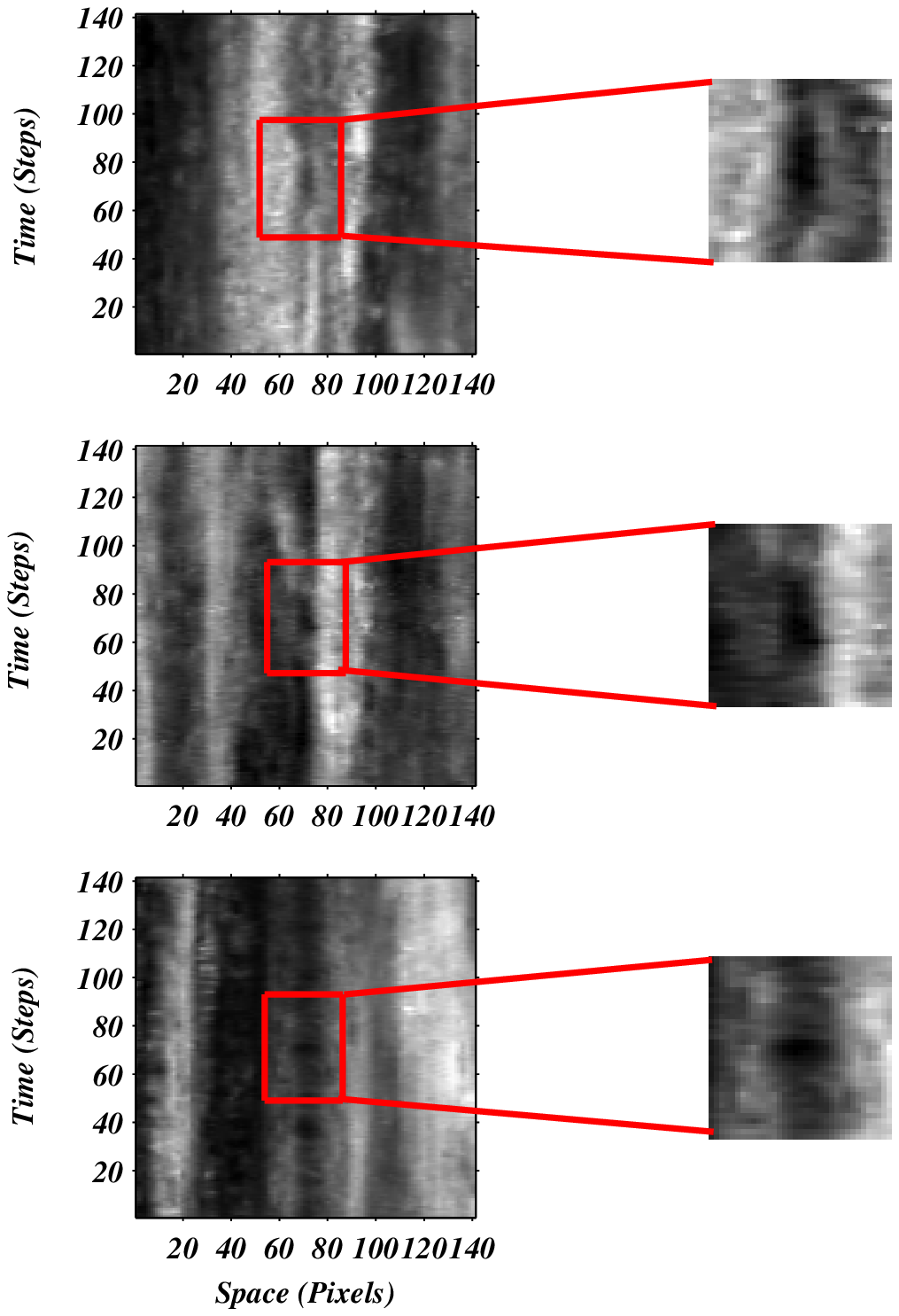}
      \caption{Three less clear events detected by SVM from AIA
        data. }
            \label{fig6}
\end{figure}

 {To make an estimate of how reliable the algorithm is we checked
400 of the detected events. We found that 30 of them are not clear events, like those shown in Figure~\ref{fig6}. We also checked for events that the algorithm missed and could not find any additional events. Therefore we may be over-estimating small events numbers by about 10\%.}

\begin{figure}
% \special{psfile=fig/fig11.eps vscale=70 hscale=70 hoffse=-30
% voffset=-210}
% \vspace{8.cm}
    \centering
\includegraphics[width=\linewidth]{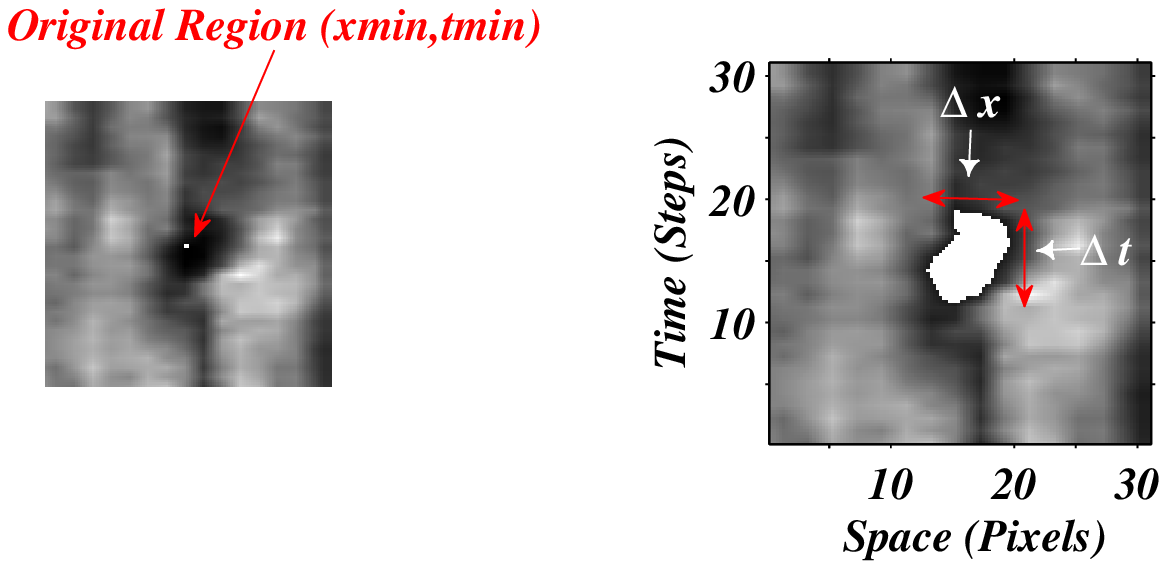}
       \caption{Demonstration of the mini-dimming extraction from a space-time slice,
The frames are 7.8 arcsec by 10.9 min from SDO/AIA.}
            \label{fig9}
\end{figure}

The size and expansion rates of the dimmings have been computed with
a region grow algorithm applied to the dimming region in the space-time
slices (Figure~\ref{fig9}).
 {First a region (2 by 2) including the
the original $x_{min}$ and $t_{min}$ is selected (left panel of
Figure \ref{fig9}). Then the region is expanded
 to include all connected pixels which fall below the specified threshold.
The width, $\Delta x$, is the maximum width of the dimming
region. The event duration time interval, $\Delta t$, is the maximum
along the time direction.}
 {Because the same event is often seen in several neighbouring $y$-slices, we first group events
together and then apply the region grow algorithm to all events in the $x$ and $y$ directions. The longest time intervals and biggest widths are taken as the event sizes.}

 {The size of the region depends on the threshold used. If it is too low then external dark areas will be included and if it is too high the region is too small.
Our automatic method uses the full width at half maximum
 of the histograms of intensities
of each space-time slice as the threshold.
 We tested this automatic level against manually chosen ones for a random 10\%\ of the slices. Here we took special care to ensure no external areas were included. The difference between the width and duration
of the two cases (automatic thresholds and
manual thresholds) are calculated. Only two out of 140 events differed by more than 2 spatial pixels (1000~km). The duration has a slightly large uncertainty: 15 of the 140 events differed by for than 2 time steps (180~s).}
% The errors, shown in Figure~\ref{error}
%in $\Delta x$ and $\Delta t$ are in the range of
%[-1300,2800]~km and [-630,450]~s, respectively.
%The averages are 516~km and 107~s.}

\begin{figure}
% \special{psfile=sunevents.eps vscale=40 hscale=36 hoffset=35
% voffset=-280}
% \vspace{8.1cm}
    \centering
\includegraphics[width=\linewidth]{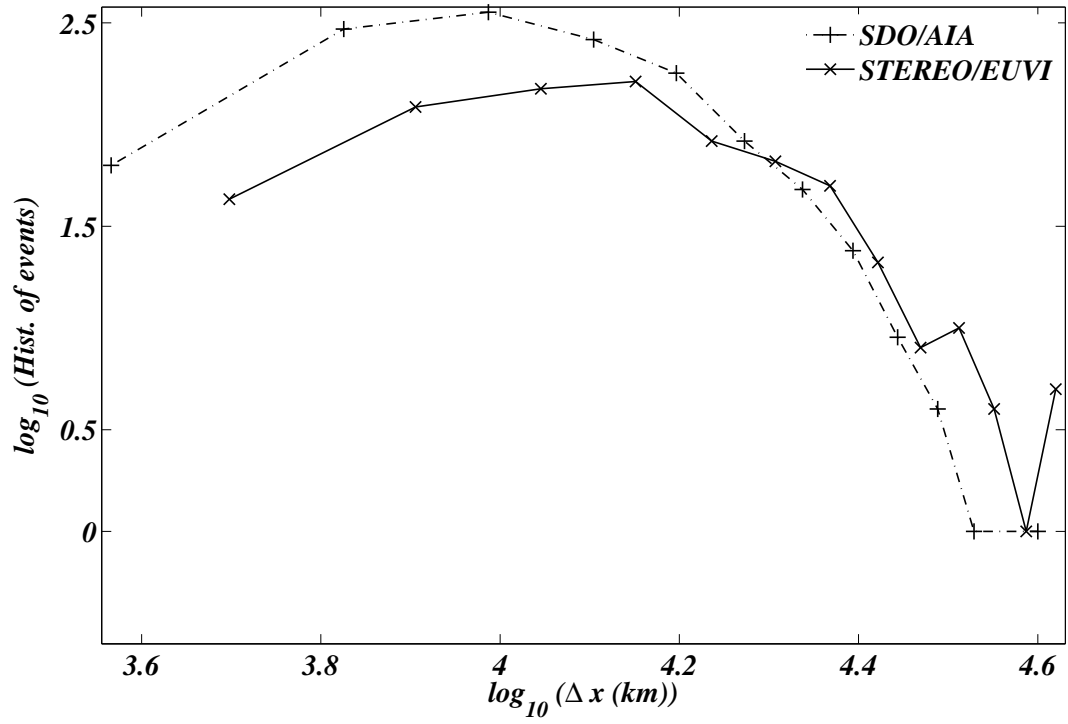}
       \caption{Histogram of  {small dimming events} detected by AIA and EUVI.}
            \label{fig10}
\end{figure}

 The histograms of the AIA and EUVI sizes are shown in
Figure~\ref{fig10}. For larger events the slopes are the same
(about -5) for both instruments. The difference between the two
curves is that AIA curve turns over at a smaller size. The size
refers to the central core dimmings not the event size so
comparison with the large-scale CME distribution (Schrijver 2010)
is not possible.

\begin{figure}
% \special{psfile=fig/fig13.eps vscale=30 hscale=30 hoffset=25
% voffset=-190}
% \vspace{6.5cm}
   \centering
\includegraphics[width=\linewidth]{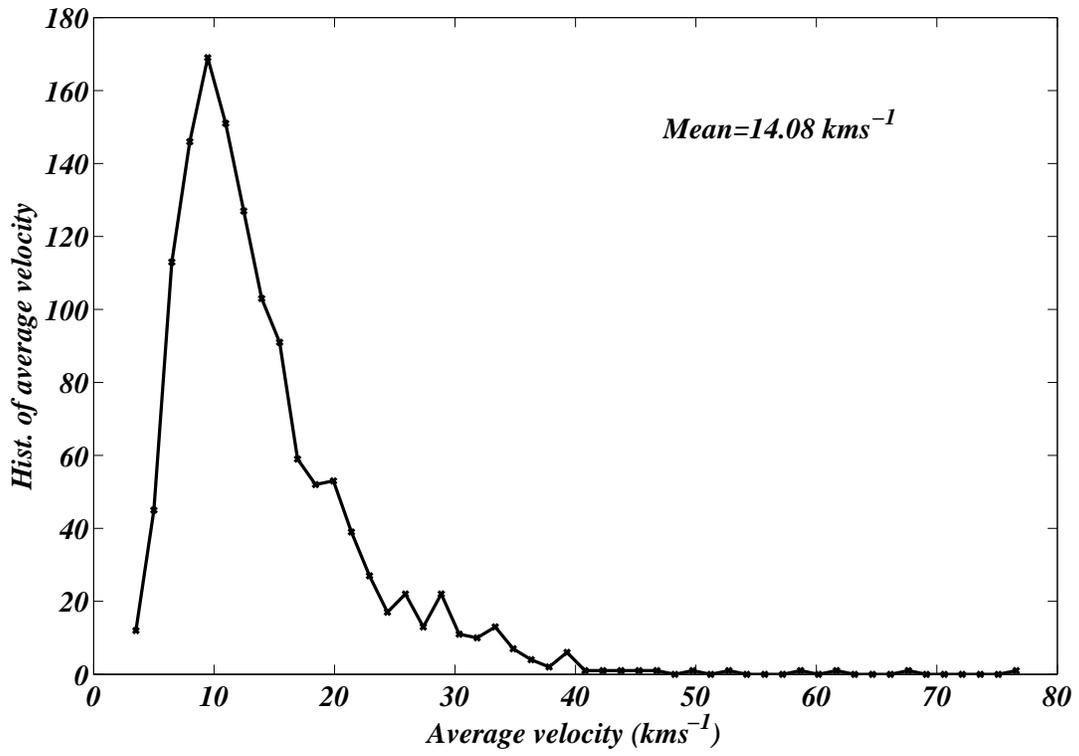}
       \caption{ The histogram of average velocities of 1325 events are plotted versus
       velocities. }
            \label{fig11}
\end{figure}

 {We compute the average velocity ($\Delta x/\Delta t$) for 1325 AIA dimming
regions. }
 The histogram of average
velocities are shown in Figure~\ref{fig11}. This histogram is
similar to the distribution of CME's speeds obtained from solar
cycle 23 (Mittal \& Narian 2009). The average velocities
concentrate in the range of 3-80~\kms\ with mean of 14~\kms.
 This is same as the average mini-filament velocity measured by Wang et al. (2000).
 {Based on the uncertainly in space and duration the velocity uncertainties are small, typically 1~\kms. Only 5\%\ have an uncertainly greater than 5~\kms.}
%The velocity uncertainties are in the range of [-6,5~km~s$^{-1}$ and the
%mean is about $0.9$~km~s$^{-1}$.}

\subsection{Conclusion}
An automatic method for the detection of  {small-scale EUV
dimmings} has been developed and tested on SDO/AIA and
STEREO/EUVI 171\AA\ images. The method exploits the fact that the
Zernike moments of events that move have a specific pattern in
space-time slices . This structure can be recognized by a Support
Vector Machine classifier (SVM). Due to the invariant
characteristics of the Zernike moments, the proposed method can
detect events even if they are small or very faint.

 {At the moment we detect motion of the dimming front but the algorithm is not able to distinguish between events due to coronal field evolution with and without eruption. Since eruptive events almost always have sudden brightening at onset, we are working on an algorithm that uses event groups to find the position and intensity of the brightest pixel at the earliest time in a group. Then the event characteristics relative to this brightening can be used to classify events as eruptive or not.}

\begin{thebibliography}{}
\bibitem{attrill1} Attrill, G. D. R. \& Wills-Davey, M. J., 2010, Sol. Phys., 262, 461.
\bibitem{asch1}Aschwanden, M. J., 2010, Sol. Phys., 262, 235.
\bibitem{Innes}Innes, D. E., Genetelli, A., Attie, R., \&
Potts, H.E. 2009, A\&A, 495, 319.
\bibitem{Innes2} Innes, D. E., McIntosh, S. W., and
 Pietarila, A. 2010, A\&A, 517, L71.
\bibitem{Bruge} Burges, C. J. C. 1998, Data Mining and Knowledge Discovery, 2,
121
\bibitem{Gunn} Gunn, S. R., 1997, Technical Report, Image Speech and Intelligent Systems Research Group, University of Southampton.
%\bibitem{Hong}Hong, J., Jiang, Y., Zheng, R., et al. 2011, ApJ, 738, L20.
\bibitem{Howard}Howard, R. A., Moses, J. D., Vourlidas, A., et al. 2008, Space Sci. Rev., 136, 67.
\bibitem{Hu1962} Hu, M. K. 1962, IRE Trans. on
Information Theory, IT-8, 179.
\bibitem{NishantNarain} Mittal, N. \& Narain, U. 2009, New Astronomy, 14, 341.
\bibitem{Podladchikova2}Podladchikova, O.V., Berghmans, D. 2005, Sol. Phys., 228, 265.
\bibitem{Podladchikova1} Podladchikova, O., Vourlidas, A.,
  Van der Linden, R. A. M., W\"{u}lser, J.-P.W, \& Patsourakos, S.
2010, ApJ, 709, 369.
\bibitem{Schrijver}Schrijver, Carolus, J. 2010, ApJ, 710, 1480.
\bibitem{wang}Wang, J., Li, W., Denker, C., et al. 2000, ApJ, 530, 1071.
%\bibitem{Zheng} Zheng, R., Jiang, Y., Hong, J., et al. 2011, ApJ, 739, L39.
\end{thebibliography}
\end{document}